\documentclass[prb,twocolumn, amsmath,amssymb]{revtex4-2}
\usepackage{graphicx,subfigure,amsmath,bm,colordvi,times}
\usepackage{comment}
\usepackage{mathtools}
\usepackage[usenames]{color}
\usepackage[dvipsnames]{xcolor}
\usepackage{simpler-wick}
\usepackage{multirow}
\usepackage{epstopdf}
\usepackage[utf8]{inputenc}
\usepackage[english]{babel}
\usepackage{amsfonts}
\usepackage{amssymb}
\usepackage{float}
\usepackage{hyperref}
\usepackage{braket}
\usepackage[normalem]{ulem}
\hypersetup{pdfborder=0 0 0,colorlinks=true,citecolor=blue,linkcolor=blue}
\usepackage{url}
\begin{document} 
\title{Beyond Random Phase Approximation in electron-hole bilayer superfluidity}
\author{Filippo Pascucci$^{1,3}$, Stefania De Palo$^{4,5}$, Sara Conti$^1$, David Neilson$^1$, Andrea Perali$^2$, Gaetano Senatore$^5$}
\affiliation{$^1$CMT Group, Department of Physics, University of Antwerp, Groenenborgerlaan 171, 2020 Antwerp, Belgium}
\affiliation{$^2$CQM Group, School of Pharmacy, University of Camerino, 62032 Camerino (MC), Italy}
\affiliation{$^3$TQC Group, Department of Physics, University of Antwerp, Universiteitsplein 1, 2610 Antwerp, Belgium}
\affiliation{$^4$
CNR-IOM Democritos, via Bonomea, 265 - 34136 Trieste, Italy}
\affiliation{$^5$Dipartimento di Fisica, Università di Trieste, Strada Costiera 11, 34151 Trieste, Italy}

\date{\today}

\begin{abstract}

We derive the normal and anomalous proper polarization functions and the screened Coulomb interactions in a two-dimensional superfluid electron-hole bilayer, including all first-order corrections beyond the Random Phase Approximation (RPA). This requires a modification of the perturbation method as first noted by Nozières  and Schrieffer \cite{Nozieres1998Chapter,Schrieffer1999Chapter}.  We discuss the physical origin and magnitude of the first-order corrections in a superfluid system with long-range Coulomb interactions. Unlike conventional superconductivity, Migdal’s theorem does not apply here, so exchange vertex corrections cannot be neglected. The screened electron-electron, hole-hole, and electron-hole interactions in the superfluid state are evaluated as functions of the carrier density. We find that at low density, the strong cancellations between the normal and anomalous components that make screening of the interactions negligible, apply not only within RPA but also with the first-order corrections included. As the density is increased, the normal-anomalous cancellation weakens and screening becomes increasingly significant. We find that the first-order corrections amplify the normal-anomalous difference but only at large momenta exchanged in the two-particle scattering, so their effect on the interactions remains modest. We conclude that the superfluid state RPA is an excellent approximation for the screening and for the effective electron-hole pairing in this superfluid system over the range of densities up to the maximum of the superfluid gap.
\end{abstract}

\maketitle
\section{Introduction}

Understanding the role of screening in strongly correlated systems remains a central problem in many-body physics \cite{Hilton1963,Mott1968,Tokura1998,Andrei2020,Strydom2023}. Coulomb-mediated electron-hole pairs systems provide a compelling platform to explore the interplay between different many-body phases and screening processes \cite{Nikolaev2004,Lozovik2012}. 

Keldysh and Kopaev \cite{Keldysh1965} proposed that bound electron-hole pairs, excitons, in semiconductors could undergo quantum condensation. The pairing would be driven by the instantaneous strong Coulomb attraction between the electrons and holes. This could lead to a superfluid phase with high critical temperature. 
In electron-hole bilayer semiconductor systems, the electrons and holes are separated into two adjacent 2D conducting layers to suppress the rapid electron-hole recombination and to stabilize long-lived excitons, hence creating conditions favorable for equilibrium exciton superfluidity \cite{Kogan1971, Lozovik1975, Lozovik1976, Shevchenko1976}. The screening here has been shown to play a central role \cite{Lozovik2012,Perali2013}. On the one hand, screening may drastically weaken the effective pairing interaction and so suppress superfluidity \cite{Kharitonov2008}. On the other hand, a large superfluid gap would block the low-energy excitations responsible for screening, and this can permit emergence of a strongly coupled superfluid \cite{Lozovik2012}.  

The screening in electron-hole bilayer system has been extensively studied within a self-consistent Random-Phase-Approximation (RPA) \cite{Sodemann2012, Perali2013, Mazloom2018} by including the zero-order (density-density) polarization functions in the screened interactions. 
The regime of validity of the RPA remains an open question in these systems.
Accurate treatment of the screening is crucial since, depending on the screening, the system may be in any of the BCS-BEC crossover regimes of the exciton superfluid  \cite{LopezRios2018,Maezono2013,DePalo2023}, or it may be a supersolid \cite{Conti2023}, or a normal dipolar crystal \cite{Astrakharchik2007}, or a normal liquid, or two coupled Wigner crystals \cite{Neilson1992,DePalo2002}. Experimentally, the screening of the electron-hole pairing attraction at low temperatures can be tuned by adjusting the electron and hole densities \cite{Perali2013, Lee2014, Zeng2020} or by selecting samples with different separations between the layers \cite{Conti2023}.

In conventional superconductors, first-order corrections in the interaction, beyond the RPA framework, introduce important phenomena including mass renormalization, energy-dependent pairing, and finite quasiparticle life-times \cite{Marsiglio2018}.  This has enabled accurate predictions of superconducting observables such as the gap and critical temperature \cite{McMillan1968, Allen1975, Pellegrini2024, Ummarino2025, Profeta2012}.

Conventional superconductors are well described by Bardeen-Cooper-Schrieffer (BCS) theory which models the pairing of electrons via an instantaneous momentum-independent contact interaction near the Fermi surface \cite{Cooper1956, Bardeen1957}.
Eliashberg theory  \cite{Eliashberg1960} extends the BCS framework by including the first-order terms from the interaction in the zero-order proper electron propagator: the electron self-energy and the vertex exchange corrections due to phonon exchange. These corrections account for the full frequency and momentum dependence of a retarded electron-phonon interaction. Under Migdal’s theorem, which is valid when the phonon energy scale is much smaller than the Fermi energy, vertex corrections can be neglected. This simplifies the theory to a self-consistent treatment of the dynamic self-energies only \cite{Migdal1958}. 

In this work, we establish the limits of validity of the RPA in a 2D exciton superfluid by evaluating for the first time all the first-order corrections to the zero-order polarization functions and determining their effect on the screened interactions. 
In contrast to conventional superconductors, Migdal's theorem does not apply here and so the vertex exchange corrections cannot be neglected. 

In Section II, we set up the formalism. Section III presents for the first time a complete and detailed derivation of the self-consistent screened interactions. Section IV evaluates the zero-order and first-order normal and anomalous polarization functions. Section V contains a detailed analysis of the polarization functions including all the first-order corrections and their effect on the screened interactions. Our conclusions appear in Section VI. The Appendices contain the details of the screened interactions and Green function calculations for the zero-order and first-order polarization terms.    

\section{Bilayer system}

The electron-hole bilayer system is made up of spatially separated parallel $n$-doped and $p$-doped conducting layers, mutually isolated by a thin insulating layer.
Each layer has a uniform neutralizing background.
For convenience, a particle-hole transformation is used to map the empty electron states in the valence band of the $p$-doped layer to a conduction band populated by positively charged holes. 
We consider only equal electron and hole densities, which is the optimal condition for large-gap and high superfluid critical temperature \cite{Pieri2007}.
The Hamiltonian for the system is:
\begin{widetext}
\begin{align}
H&=H_0+H_1, \label{ham} \\
H_0&=\sum_{\mathbf{k},\sigma} \epsilon_{\mathbf{k},\sigma} \left(c^\dag_{\mathbf{k},\sigma}c_{\mathbf{k},\sigma}+d^\dag_{\mathbf{k},\sigma}d_{\mathbf{k},\sigma}\right) \label{H0},  \\ 
H_1&=\frac{1}{A} \sum_{\substack{\mathbf{k},\mathbf{k'},\mathbf{q}\neq 0,\\\sigma,\sigma'}}
\left[\frac{V^{S}_{\mathbf{q}}}{2} 
\left(c^\dag_{\mathbf{k}+\mathbf{q},\sigma}c^\dag_{\mathbf{k'}-\mathbf{q},\sigma'}c_{\mathbf{k'},\sigma'}c_{\mathbf{k},\sigma}
+d^\dag_{\mathbf{k}+\mathbf{q},\sigma}d^\dag_{\mathbf{k'}-\mathbf{q},\sigma'}d_{\mathbf{k'},\sigma'}d_{\mathbf{k},\sigma} \right)+
V^{D}_{\mathbf{q}}c^\dag_{\mathbf{k}+\mathbf{q},\sigma}d^\dag_{\mathbf{k'}-\mathbf{q},\sigma'}d_{\mathbf{k'},\sigma'}c_{\mathbf{k},\sigma} \right] \ ,
\label{ham1}
\end{align}
\end{widetext}
where $c_{\mathbf{k},\sigma}$($d_{\mathbf{k},\sigma}$) and $c^\dag_{\mathbf{k},\sigma}$($d^\dag_{\mathbf{k},\sigma}$) are the electron (hole) annihilation and creation operators in the Schrodinger representation,  with momentum $\mathbf{k}$ and  $\sigma=(v,s_z)$ the valley-spin index or flavor. 
We consider equal electron and hole effective masses $m^*$, with energy dispersion $\sigma$-independent $\epsilon_{\mathbf{k}}=\frac{\hbar^2|\mathbf{k}|^2}{2m^*}-\mu_s$ and single-particle chemical potential $\mu_s$.
$V^{S}_{\mathbf{q}}=V^{ee}_{\mathbf{q}}= V^{hh}_{\mathbf{q}}= \frac{2\pi e^2}{4\pi\epsilon |\mathbf{q}|}$ and $ V^D_{\mathbf{q}}=V^{eh}_{\mathbf{q}}=V^{he}_{\mathbf{q}}=-V^S_{\mathbf{q}}\textrm{e}^{-|\mathbf{q}|d}$ are the bare Coulomb instantaneous interactions with $d$ the thickness of the insulating barrier separating the layers and $\epsilon$ its dielectric constant. $A$ is the area of each layer.
 
 The extension of perturbation methods, i.e., S-matrix formalism, to superfluids requires \cite{Nozieres1998Chapter,Schrieffer1999Chapter} to start from an unperturbed state that is already superfluid/superconductor. This  can be achieved by introducing an auxiliary linear hamiltonian $H_{lin}$, so that the unperturbed hamiltonian becomes 
\begin{equation}
    H_0'=H_0+H_{lin} \ ,
\end{equation}
and the perturbing hamiltonian  
\begin{equation}
H_1'=H_1-H_{lin}\ .
\end{equation} 
Clearly
\begin{equation}
H_0'+H_1'=H\ .
\end{equation}
A suitable choice of $H_{lin}$ is  
\begin{align}    
H_{lin}&=\sum_{\mathbf{k},\sigma}\left[\chi_{\mathbf{k},\sigma}\left( c^\dag_{\mathbf{k},\sigma}c_{\mathbf{k},\sigma}+d^\dag_{\mathbf{k},\sigma}d_{\mathbf{k},\sigma}\right)\right.\\
   &\left.+\left(\Delta_{\mathbf{k}, \sigma} c^\dag_{\mathbf{k},\sigma}d^\dag_{\mathbf{-k},\sigma}+\Delta_{\mathbf{k}, \sigma}^* d_{\mathbf{-k},\sigma}c_{\mathbf{k},\sigma}  \right)\right] ,
    \label{HintBCS}
\end{align}
yielding 
\begin{align}    
H_0'&=\sum_{\mathbf{k},\sigma}\left[\xi_{\mathbf{k},\sigma}\left( c^\dag_{\mathbf{k},\sigma}c_{\mathbf{k},\sigma}+d^\dag_{\mathbf{k},\sigma}d_{\mathbf{k},\sigma}\right)\right. \nonumber
 \\ 
   &\left.+\left(\Delta_{\mathbf{k}, \sigma} c^\dag_{\mathbf{k},\sigma}d^\dag_{\mathbf{-k},\sigma}+\Delta_{\mathbf{k}, \sigma}^*d_{\mathbf{-k},\sigma}c_{\mathbf{k},\sigma}  \right)\right] ,
    \label{HintBCS1}
\end{align}
with
\begin{equation}
\xi_{\mathbf{k},\sigma}=\epsilon_{\mathbf{k},\sigma} +\chi_{\mathbf{k},\sigma}.
\end{equation}
Results will makes sense only if they are independent of the exact choice of $H_0'$ \cite{Nozieres1998Chapter}. The parameters in $H_0'$ (i.e., the sets of $\chi_{\mathbf{k},\sigma}$ and $\Delta_{\mathbf{k}, \sigma}$) remain to be determined at this stage.  
One possibility is to determine them by comparing  Eq.\ \eqref{HintBCS1} with the mean-field linearized version of Eq.\ \eqref{ham}, term by term.
If we resort to a state of the BCS form, 
\begin{equation}
      \ket{\Psi_{BCS}} =\prod_{\mathbf{k},\sigma} \left(u_\mathbf{k}+v_\mathbf{k}c_{\mathbf{k},\sigma}^\dag d_{-\mathbf{k},\sigma}^\dag\right)\ket{0} \label{Psi0}\ ,\ 
\end{equation}
and minimize $\bra {\Psi_{BCS}} H_0' \ket{\Psi_{BCS}}$ with respect to  the   Bogoliubov amplitudes $u_\mathbf{k}$ and $v_\mathbf{k}$
we obtain: 
\begin{equation}
u_\mathbf{k}^2=\frac{1}{2}\left(1+\frac{\xi_{\mathbf{k}}}{E_{\mathbf{k}}}\right), 
v_\mathbf{k}^2=\frac{1}{2}\left(1-\frac{\xi_{\mathbf{k}}}{E_{\mathbf{k}}}\right),   
E_{\mathbf{k}}=\sqrt{\xi_k^2+\Delta_\mathbf{k}^2},
\end{equation}
as well as  
\begin{equation} 
\chi_{\mathbf{k},\sigma}=\chi_\mathbf{k}=-\frac{1}{A}\sum_{\mathbf{k'}}V^S_{\mathbf{k-k'}}v_{\mathbf{k'}}^2.
\label{eq:hf_nrg}
\end{equation}
Evidently, $\chi_{\mathbf{k},\sigma}$ is the self-consistent Hartree-Fock intralayer energy. 
Manipulation of the relation between $\Delta_{\mathbf{k}, \sigma}$ in  Eq.\ \eqref{HintBCS1} and  $ \tilde{\Delta}_{\mathbf{k}, \sigma}=\bra {\Psi_{BCS}} d_{\mathbf{-k},\sigma}c_{\mathbf{k},\sigma} \ket{\Psi_{BCS}}$  in the mean-field linearized version of Eq.\ \eqref{ham} finally
 yields to
 \begin{equation}
  \Delta_{\mathbf{k},\sigma}=  \Delta_\mathbf{k}=-\frac{1}{A}\sum_{\mathbf{k'}}V^D_{\mathbf{k}-\mathbf{k}'}\frac{\Delta_{\mathbf{k'}}}{2E_{\mathbf{k'}}}\ ,
    \label{eq:gap_eq}
\end{equation}
the self-consistent gap energy. In summary, the self-consistent solution of the following equations
\begin{align}
\xi_{\mathbf{k}}& =\epsilon_{\mathbf{k}} -\frac{1}{A}\sum_{\mathbf{k'}}V^S_{\mathbf{k-k'}}v_{\mathbf{k'}}^2,\\
 \Delta_\mathbf{k}&=-\frac{1}{A}\sum_{\mathbf{k'}}V^D_{\mathbf{k}-\mathbf{k}'}\frac{\Delta_{\mathbf{k'}}}{2E_{\mathbf{k'}}},\\
 E_{\mathbf{k}}&=\sqrt{\xi_\mathbf{k}^2+\Delta_\mathbf{k}^2},
\end{align}
at given $\mu_s$, (recall $\epsilon_{\mathbf{k}}=\frac{\hbar^2|\mathbf{k}|^2}{2m^*}-\mu_s$), provides 
the gap energy  $ \Delta_\mathbf{k}$ and the in-layer density
\begin{equation}
n=\frac{1}{A}\sum_{\mathbf{k},\sigma}v_\mathbf{k}^2 =\frac{g}{A}\sum_{\mathbf{k}}v_\mathbf{k}^2,
\end{equation}
where  $g=g_sg_v$ is the spin-valley degeneracy prefactor and  $n=n_e=n_h$, as we have a symmetric e-h bilayer.

\section{ Screened interaction in the exciton superfluid state}
To obtain test charge-test charge screened interactions \cite{Giuliani2005} in a one-component system, one may resort to the  proper polarization 
$\Pi^*(\mathbf{q},\omega)$ (see, e.g., Eq. 9.40 in \cite{Fetter1971}) to write
\begin{equation}
    W(\mathbf{q},\omega)=V(\mathbf{q})+V(\mathbf{q}) \Pi^*(\mathbf{q},\omega)W(\mathbf{q},\omega) \ ,
    \label{Dyseq1}
\end{equation}
where $V(\mathbf{q})$ is the bare interaction and  $W(\mathbf{q},\omega)$ the screened interaction. $V(\mathbf{q}), W\mathbf{q},\omega), \Pi^*(\mathbf{q},\omega)$ are scalars in a one-component system. In a two-component system, such as the electron-hole  symmetric bilayer considered here, $V(\mathbf{q}), W(\mathbf{q},\omega), \Pi^*(\mathbf{q},\omega)$ become symmetric 2x2 square matrices. 
The bare interaction matrix is
\begin{align}
    V(\mathbf{q})=
    \begin{pmatrix}
       V^S_\mathbf{q} & V^D_\mathbf{q} \\
       V^D_\mathbf{q} & V^S_\mathbf{q}
    \end{pmatrix}
    \ ,
    \label{VmatrixSF2}
\end{align} 
and the proper polarization matrix 
\begin{align}
    \Pi^*(\mathbf{q},\omega)=
    \begin{pmatrix}
       \Pi^{N}(\mathbf{q},\omega) & \Pi^{A}(\mathbf{q},\omega) \\
       \Pi^{A}(\mathbf{q},\omega) & \Pi^{N}(\mathbf{q},\omega) 
    \end{pmatrix} \ ,
    \label{PmatrixSF2}
\end{align}
where $\Pi^{N}(\mathbf{q},\omega)=\Pi^{ee}(\mathbf{q},\omega)=\Pi^{hh}(\mathbf{q},\omega)$ is the electron-electron normal proper polarization function and $\Pi^{A}(\mathbf{q},\omega)=\Pi^{eh}(\mathbf{q},\omega)=\Pi^{he}(\mathbf{q},\omega)$ is the electron-hole anomalous proper polarization function.
The screened interaction matrix is obtained  from the matrix version of Eq. \eqref{Dyseq1} as
\begin{align}
     W(\mathbf{q},\omega)=
    \begin{pmatrix}
        V^S_{sc}(\mathbf{q},\omega) & V^D_{sc}(\mathbf{q},\omega) \\
        V^D_{sc}(\mathbf{q},\omega) & V^S_{sc}(\mathbf{q},\omega)
    \end{pmatrix}\ .
   \label{WmatrixSF}
\end{align}

In particular, in the static limit $\omega\rightarrow0$ one finds \cite{Pascucci2024}
\begin{align}
 V^S_{sc}(\mathbf{q})=\frac{V^S_\mathbf{q}-\Pi^N(\mathbf{q})\mathcal{A}_\mathbf{q}}{1\!-\!2(\Pi^N(\mathbf{q})V^{S}_\mathbf{q}+\Pi^A(\mathbf{q})V^{D}_\mathbf{q})+\mathcal{B}_{\mathbf{q}}\mathcal{A}_\mathbf{q}},
 \label{eq-uno}
\end{align}
and
\begin{align}
 V^D_{sc}(\mathbf{q})=\frac{V^D_\mathbf{q}+\Pi^A(\mathbf{q})\mathcal{A}_\mathbf{q}}{1\!-\!2(\Pi^N(\mathbf{q})V^{S}_\mathbf{q}+\Pi^A(\mathbf{q})V^{D}_\mathbf{q})+\mathcal{B}_{\mathbf{q}}\mathcal{A}_\mathbf{q}},
 \label{eq-due}
\end{align}
with 
\begin{equation}
\mathcal{A}_\mathbf{q}\!=\!\left(V^{S}_\mathbf{q}\right)^2-\left(V^{D}_\mathbf{q}\right)^2, \,\,\mathcal{B}_{\mathbf{q}}\!=\!\left(\Pi^N(\mathbf{q})\right)^2-\left(\Pi^A(\mathbf{q})\right)^2. \label{eq-tre}
\end{equation}
Eqs.\ \eqref{eq-uno}-\eqref{eq-tre} provide the screened interactions corresponding to a choice of the proper polarization matrix $\Pi^*(\mathbf{q})$. In the following we shall compute, for selected values of density,  the screened interactions obtained (i) from the zero-order $\Pi^*(\mathbf{q})$ (RPA) and (ii) from the $\Pi^*(\mathbf{q})$ including zero- and first-order. See below.
Details about the derivation of Eqs. \eqref{eq-uno}-\eqref{eq-tre}  are found in the Appendix \ref{App:screened}.

\section{Polarization functions}
The proper polarization functions are defined as :
\begin{align}
    \Pi^{N}(\mathbf{q},\omega)&=-\frac{i}{\hbar A}\ \lim_{\eta\rightarrow0}\int_{-\infty}^{\infty} d\tau e^{i(\omega\tau+i\eta|\tau|)}\langle T[\rho_e \rho_e S]\rangle_c \, , \label{PN*}\\
    \Pi^{A}(\mathbf{q},\omega)&=-\frac{i}{\hbar A }\ \lim_{\eta\rightarrow0}\int_{-\infty}^{\infty} d\tau e^{i(\omega\tau+i\eta|\tau|)}\langle T[\rho_e \rho_h S]\rangle_c \, \label{PA*} .
\end{align}
Above,
\begin{equation}
  \langle T[\rho_\lambda \rho_{\lambda'}S]\rangle_c=\frac{\langle\Psi_{BCS}|T[\rho_\lambda(\mathbf{q},\tau)\rho_{\lambda'}(-\mathbf{q},0)S]|\Psi_{BCS}\rangle}{\langle\Psi_{BCS}|S|\Psi_{BCS}\rangle} 
\end{equation}
is the sum of the expectation values of the connected diagrams only \cite{Fetter1971}, evaluated on the excitonic BCS ground state $\Psi_{BCS}$ (Eq.\ \eqref{Psi0}). 
 $\rho_e(\mathbf{q},\tau)$ and $\rho_h(\mathbf{q},\tau)$ are the density fluctuation fields, 
\begin{align}
    \rho_e(\mathbf{q},\tau)&=\sum_\mathbf{k,\sigma} c^\dag_{\mathbf{k}+\mathbf{q},\sigma}(\tau)c_{\mathbf{k},\sigma}(\tau)\ ,\\    
    \rho_h(\mathbf{q},\tau)&=\sum_\mathbf{k,\sigma} d^\dag_{\mathbf{k}+\mathbf{q},\sigma}(\tau)d_{\mathbf{k},\sigma}(\tau)   \ .
\end{align}
The expanded scattering matrix $S$ reads \cite{Dyson1949}:
\begin{equation}
    S=\!T\!\left[e^{-\frac{i}{\hbar}\int d\tau' H_{1}'(\tau')}\right]\!\simeq\! T\left[1-\frac{i}{\hbar}\int d\tau' H_{1}'(\tau')+...\right] , \label{Scatmat}
\end{equation}
where we truncate the expansion to first order in $H_{1}'$.

\subsection{Zero-order}
The zero-order of the scattering matrix expansion in Eqs.\ \eqref{PN*}-\eqref{PA*}, $S=1$, is the response
function in the mean-field approximation.
\begin{align}
    \Pi_0^N(\mathbf{q},\omega)=-\frac{i}{\hbar A }\ \lim_{\eta\rightarrow0}\int d\tau e^{i(\omega\tau+i\eta|\tau|)}\langle T[\rho_e\rho_e]\rangle_c \label{PiSRPA} \ ,\\
        \Pi_0^A(\mathbf{q},\omega)=-\frac{i}{\hbar A}\ \lim_{\eta\rightarrow0}\int d\tau e^{i(\omega\tau+i\eta|\tau|)}\langle T[\rho_e\rho_h]\rangle_c \label{PiDRPA} \ .
\end{align}
We introduce the normal $G_{\sigma}(\mathbf{k},\tau)$ and anomalous $F_{\sigma}(\mathbf{k},\tau)$ Green functions (see Appendix \ref{App:0order}),
\begin{align}
    iG_{\sigma}(\mathbf{k},\tau)&=\langle T[c_{\mathbf{k},\sigma}(\tau) c^{\dag}_{\mathbf{k},\sigma}(0)]\rangle\nonumber\\
    &=\theta(\tau)u_{\mathbf{k}}^2e^{-iE_{\mathbf{k}}\tau/\hbar}- \theta(-\tau)v_{\mathbf{k}}^2e^{iE_{\mathbf{k}}\tau/\hbar} \ , \label{normalG}\\
    \nonumber\\
      iF_{\sigma}(\mathbf{k},\tau)&=\langle T[c_{\mathbf{k},\sigma}(\tau)d_{-\mathbf{k},\sigma}(0)]\rangle\nonumber\\
      &=v_{\mathbf{k}}u_{\mathbf{k}}\left[\theta(\tau)e^{-iE_{\mathbf{k}}\tau/\hbar}+\theta(-\tau)e^{iE_{\mathbf{k}}\tau/\hbar}\right]\,  , \label{anomF} 
\end{align}
so that:
\begin{align}
    \langle T[\rho_e(\mathbf{q},\tau)\rho_e(-\mathbf{q},0)]\rangle_c&= g\sum_{\mathbf{k}}G_{\mathbf{k}+\mathbf{q}}(-\tau)G_\mathbf{k}(\tau), \label{Tee} \\
    \langle T[\rho_e(\mathbf{q},\tau)\rho_h(-\mathbf{q},0)] \rangle_c&=-g\sum_{\mathbf{k}}F^{*}_{\mathbf{k}+\mathbf{q}}(\tau)F_\mathbf{k}(\tau) \ \label{Teh} .
\end{align}
In the static limit $\omega\rightarrow0$, 
\begin{align}
\Pi_0^N(\mathbf{q})&\!=\!-\frac{ig}{\hbar A}\!\sum_{\mathbf{k}}\lim_{\eta,\omega\rightarrow0}\!\!\int\!\! d\tau e^{i(\omega\tau+i\eta|\tau|)}G_{\mathbf{k}+\mathbf{q}}(-\tau)G_\mathbf{k}(\tau)\nonumber\\
&=-\frac{g}{A} \sum_\mathbf{k}\frac{u_\mathbf{k}^2v_{\mathbf{k}+\mathbf{q}}^2+v_\mathbf{k}^2u_{\mathbf{k}+\mathbf{q}}^2}{E_{\mathbf{k}}+E_{\mathbf{k}+\mathbf{q}}} \label{P0N} \ ,\\
        \Pi_0^A(\mathbf{q})&\!=\!\frac{ig}{\hbar A}\!\sum_{\mathbf{k}}\! \lim_{\eta,\omega\rightarrow0}\!\!\int\!\! d\tau e^{i(\omega\tau+i\eta|\tau|)} F^{*}_{\mathbf{k}+\mathbf{q}}(\tau)F_\mathbf{k}(-\tau)\nonumber\\
        &= -\frac{g}{2A} \sum_\mathbf{k}\frac{\Delta_\mathbf{k}\Delta_\mathbf{k+q}}{E_\mathbf{k}E_\mathbf{k+q}\left(E_{\mathbf{k}}+E_{\mathbf{k}+\mathbf{q}}\right)}\  \label{P0A} .
\end{align}
These zero-order polarization functions consist of normal and anomalous single-particle propagators, which include the effects of both electron-electron, hole-hole and electron-hole interactions at the mean-field level. When $\Delta_\mathbf{k}=0$, we recover the polarizability of an electron gas with exchange.
Substituting the zero-order static polarizabilities from Eqs.\ \eqref{P0N}-\eqref{P0A} into Eqs.\ \eqref{eq-uno}-\eqref{eq-tre}, one obtains the static RPA screened intralayer and interlayer interactions. \citep{Conti2019}.

\subsection{First-order Corrections}
Looking at the first-order terms of the scattering matrix (Eqs.\ \eqref{Scatmat}), the normal and anomalous polarization functions in Eqs.\ \eqref{PN*}-\eqref{PA*} become:
\begin{align}
    \Pi^{N}(\mathbf{q},\omega)&=\Pi_0^N(\mathbf{q},\omega)+\Pi_{1}^{N}(\mathbf{q},\omega)\ ,\\
    \Pi^{A}(\mathbf{q},\omega)&=\Pi_0^A(\mathbf{q},\omega)+\Pi_{1}^{A}(\mathbf{q},\omega)\ ,
    \end{align}
where
\begin{small}
\begin{align}
    \Pi_{1}^{N}(\mathbf{q},\omega)&=\!-\frac{g}{2\hbar^2A}\! \lim_{\eta\rightarrow0}\! \iint \!\!\!d\tau d\tau' e^{i(\omega\tau+i\eta|\tau|)}\langle T[\rho_e \rho_eH_1'(\tau')]\rangle_c \ , \label{P1*N}\\
    \Pi_{1}^{A}(\mathbf{q},\omega)&=\!-\frac{g}{2\hbar^2A} \!\lim_{\eta\rightarrow0}\! \iint\!\!\! d\tau d\tau' e^{i(\omega\tau+i\eta|\tau|)}\langle T[\rho_e \rho_hH_1'(\tau')]\rangle_c \ \label{P1*A} .
\end{align}
    \end{small}

In Appendix\ \ref{App:1order}, we evaluate the first-order corrections using the Wick theorem for time-ordered products. From the effective interaction Hamiltonian $H_1'$, the resulting connected diagrams are: (i) equal-time contractions with zero exchanged momentum (Hartree or direct terms); and (ii) non-equal-time contractions, the exchange (EX) terms.
All the Hartree terms cancel between each other due to the charge neutrality of the system.
During the calculation, all self-energy contributions cancel exactly, as shown in Appendix\ \ref{App:1order}. This cancellation originates from the structure of the chosen ground state $H_0'$, whose self-energy terms, included in linearized mean-field term $H_{lin}$, precisely compensated by those in $H_1$. The mean-field ground-state hamiltonian $H_0'$ already includes the equal-time electron–electron and electron–hole self-energy contributions through the Hartree-Fock term $\chi_\mathbf{k}$ and the BCS gap $\Delta_\mathbf{k}$. The normal and anomalous Green’s functions, Eqs.\ \eqref{normalG}–\eqref{anomF}, therefore describe quasiparticles in the superfluid phase whose propagators are already dressed by these static self-energy contributions. Including additional self-energy diagrams at first-order would thus result in double counting.
This behavior contrasts with the Migdal–Eliashberg theory, where the nature of the retarded electron–phonon interaction requires the first-order polarization functions to be evaluated using the Green’s functions of a non-interacting electron system \cite{Mahan2000, Zappacosta2025}. Consequently, in conventional superconductors the self-energy corrections are present, while exchange corrections are suppressed by Migdal’s theorem.

In the present exciton bilayer case, only the exchange terms from the electron-electron, hole-hole and electron-hole interactions remain in the first-order normal $\Pi_{1}^{N}=\Pi_{1}^{N,EX}$ and anomalous $\Pi_{1}^{A}=\Pi_{1}^{A,EX}$. Thus:
\begin{align}
\Pi_1^{N,EX}&=\Pi_{1,ee}^{N,EX}+\Pi_{1,hh}^{N,EX}+\Pi_{1,eh}^{N,EX} \\
    \Pi_1^{A,EX}&=\Pi_{1,ee}^{A,EX}+\Pi_{1,hh}^{A,EX}+\Pi_{1,eh}^{A,EX}
\end{align}
In the static limit, $\omega$$\rightarrow$$0$,  each term becomes (see Appendix \ref{App:1order}):
\begin{small}
\begin{align}
   \! \! \Pi^{N,EX}_{1,ee}\!(\mathbf{q})\!&=\!-\frac{g}{A^2}\!\sum_{\mathbf{k},\mathbf{p}}\!\frac{(u_{\mathbf{p'}}^2v_{\mathbf{p}}^2\!+\!u_{\mathbf{p}}^2v_{\mathbf{p'}}^2)(u_{\mathbf{k'}}^2v_{\mathbf{k}}^2\!+\!u_{\mathbf{k}}^2v_{\mathbf{k'}}^2)}{\left(E_\mathbf{p}+E_{\mathbf{p'}}\right)\left(E_\mathbf{k}+E_{\mathbf{k'}}\right)}V^{S}_{\mathbf{k}\!-\!\mathbf{p}} \label{P1eeNEX}
\end{align}
\end{small}

\begin{small}
\begin{align}
     \Pi_{1,hh}^{N,EX}(\mathbf{q})\!=\!-\frac{g}{2A^2}\!\sum_{\mathbf{k},\mathbf{p}}\!\frac{\Delta_\mathbf{k}\Delta_\mathbf{k'}\Delta_\mathbf{p}\Delta_\mathbf{p'}}{E_\mathbf{k}\!E_\mathbf{k'}\!E_\mathbf{p}\!E_\mathbf{p'}\!(E_{\mathbf{k}}+E_{\mathbf{k'}})\!(E_{\mathbf{p}}+E_{\mathbf{p'}})}V^S_{\mathbf{k}-\mathbf{p}}  \label{P1hhNEX}
\end{align}
\end{small}

\begin{small}
\begin{align}
    \Pi^{N,EX}_{1,eh}\!(\mathbf{q})\!=\!\frac{g}{2A^2}\!\sum_{\mathbf{k},\mathbf{p}}\!\frac{\Delta_\mathbf{k}\Delta_\mathbf{p}(v_{\mathbf{p'}}^2\!-\!u_{\mathbf{p'}}^2)(v_{\mathbf{k'}}^2\!-\!u_{\mathbf{k'}}^2)}{E_\mathbf{k}E_\mathbf{p}(E_{\mathbf{p}}+E_{\mathbf{p'}})(E_{k}+E_{\mathbf{k'}})}V^D_{\mathbf{k}-\mathbf{p}}\label{P1ehNEX}
\end{align}
\end{small}

\begin{small}
\begin{align}
    \Pi^{A,EX}_{1,ee}\!(\mathbf{q})\!&=\!\Pi^{A,EX}_{1,hh}\!(\mathbf{q})\nonumber\\
    &=\!-\frac{g}{2A^2}\!\sum_{\mathbf{k},\mathbf{p}}\!\frac{\Delta_\mathbf{p}\Delta_\mathbf{p'}(v_\mathbf{k}^2u_{\mathbf{k'}}^2\!+\!u_\mathbf{k}^2v_{\mathbf{k'}}^2)\!}{E_\mathbf{p}E_\mathbf{p'}(E_{\mathbf{k}}+E_{\mathbf{k'}})(E_{\mathbf{p}}+E_{\mathbf{p'}})}V^S_{\mathbf{k}-\mathbf{p}} \label{P1eeAEX}
\end{align}
\end{small}

\begin{small}
\begin{align}
  \!\!  \Pi^{A,EX}_{1,eh}\!(\mathbf{q})\!=\!-\frac{g}{2A^2}\!\sum_{\mathbf{k},\mathbf{p}}\!\frac{\Delta_\mathbf{k}\Delta_\mathbf{p'}(v_{\mathbf{k'}}^2\!-\!u_{\mathbf{k'}}^2)\!(u_{\mathbf{p}}^2\!-\!v_{\mathbf{p}}^2)}{E_\mathbf{k}E_\mathbf{p'}(E_{\mathbf{k}}+E_{\mathbf{k'}})(E_{\mathbf{p}}+E_{\mathbf{p'}})}V^{D}_{\mathbf{k}-\mathbf{p}} \label{P1ehAEX}
\end{align}
    \end{small}
with $\mathbf{p'}=\mathbf{p}+\mathbf{q}$ and $\mathbf{k'}=\mathbf{k}+\mathbf{q}$.
Note that when $\Delta=0$, we recover the case of normal electron-hole liquid with exchange, and only the normal electron-electron polarization function (Eq.\ \eqref{P1eeNEX}) survives.

We evaluate the screened electron-electron interaction $V_\mathbf{q}^{S,tot}$ and the screened electron-hole interaction $V_\mathbf{q}^{D,tot}$ with the first-order corrections included by replacing in Eqs.\ \eqref{eq-uno}-\eqref{eq-due}, 
$\Pi_0^N(\mathbf{q})$ with $\Pi_\textrm{tot}^N(\mathbf{q})=\Pi_0^N(\mathbf{q})+\Pi_1^{N,EX}(\mathbf{q})$ and $\Pi_0^A(\mathbf{q})$ with $\Pi_\textrm{tot}^A(\mathbf{q})\!=\!\Pi_0^A(\mathbf{q})+\Pi_1^{A,EX}(\mathbf{q})$.

\section{Results}
We use effective mass $m^*=0.04m_e$, dielectric constant $\epsilon=2\epsilon_0$ with $\varepsilon_0$ the vacuum dielectric constant, spin-valley degeneracy $g_sg_v=2$, and interlayer distance $d=0.2$ $a_B^*$ with $a_B^*=5.3$nm the effective Bohr radius as in Ref.\ \cite{LopezRios2018}. 

\begin{figure}[t]
    \centering
    \includegraphics[trim=0.0cm 0.0cm 0.0cm 0.0cm, clip=true, width=0.5\textwidth]{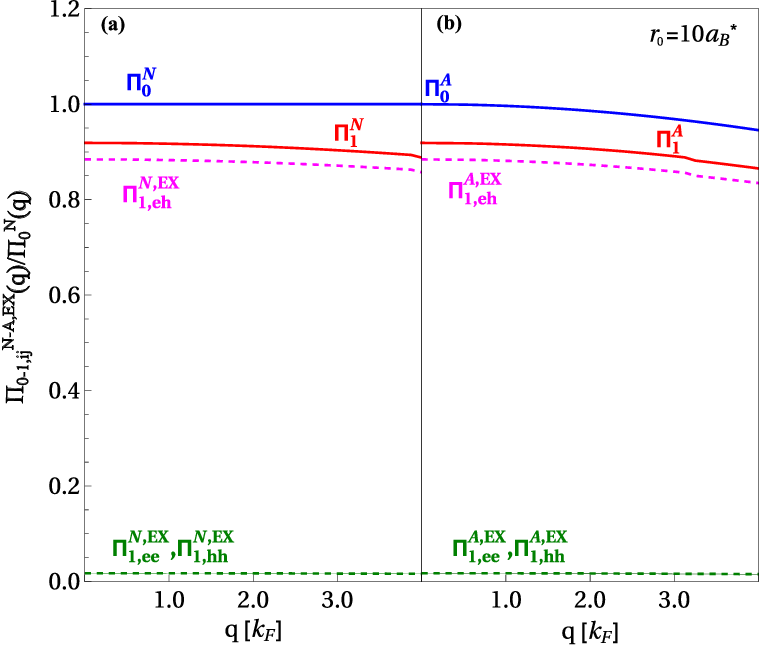}  
    \caption{(a) Normal zero-order polarization function $\Pi_0^N$ (Eq.\ \eqref{P0N}, blue solid line). Normal first-order polarization function $\Pi_1^N$ (red solid line) with components: $\Pi_{1,ee}^{N,EX}$, $\Pi_{1,hh}^{N,EX}$ (Eq.\ \eqref{P1eeNEX}-\eqref{P1hhNEX}, green dashed line), $\Pi_{1,eh}^{N,EX}$ (Eq.\ \eqref{P1ehNEX}, magenta dashed line). (b) Anomalous zero-order polarization function $\Pi_0^A$ (Eq.\ \eqref{P0A}, blue solid line). Anomalous first-order polarization function $\Pi_1^A$ (red solid line) with components: $\Pi_{1,ee}^{A,EX}$, $\Pi_{1,hh}^{A,EX}$ (Eq.\ \eqref{P1eeAEX}, green dashed line), $\Pi_{1,eh}^{A,EX}$ (Eq.\ \eqref{P1ehAEX}, magenta dashed line). Interparticle distance $r_0=10.0a_B^*$. All the curves are in unit of $\Pi_0^N(\mathbf{q})$.}
 \label{Fig_1}
\end{figure}

Figure \ref{Fig_1} shows the zero-order and first-order contributions to the normal and anomalous polarizations in unit of $\Pi_0^N(\mathbf{q})$ as a function of the momentum $\mathbf{q}$. The interparticle distance is $r_0=10a_B^*$, corresponding to a very low density, with $n=1/\pi r_0^2$.

The exchange terms have the same sign of $\Pi_0^N<0$, because they introduce an interaction line between the particle and the antiparticle propagators within the $\Pi_0$ diagram. 
This imposes a correlation that suppresses the system's ability to rearrange its charge density, leading to a negative correction to the polarization function. 

From Fig.\ \ref{Fig_1} we see that the first-order corrections of both the normal $\Pi_1^N$ and anomalous $\Pi_1^A$ polarization functions are dominated by the exchange electron-hole interaction channels $\Pi_{1,eh}^{N,EX}$ and $\Pi_{1,eh}^{A,EX}$. They are of the same order of the zero-order terms. As shown in Eqs.\ \eqref{eq-uno}–\eqref{eq-due}, the deviation of the screened interactions from the bare ones depends on the magnitude of the products $\Pi^NV^S$, as well as on their mutual cancellation. In the low-density regime, the zero-order terms $\Pi_0^NV^S$ and $\Pi_0^AV^D$ are very small, so the effect of density fluctuations on the screened interactions at zero-order is negligible.
Consequently, the first-order corrections are minimal, and Fig.\ \ref{Fig_1} shows that there is an almost complete cancellation between the normal and anomalous contributions.
The normal electron-electron (hole-hole) exchange interaction channels $\Pi_{1,ee}^{N,EX}$ ($\Pi_{1,hh}^{N,EX}$) and $\Pi_{1,ee}^{A,EX}$ ($\Pi_{1,hh}^{A,EX}$) are negligible since at such a low density most of the electrons and holes are in the condensate.

In Fig.\ \ref{Fig_1}(a), $\Pi_{1,ee}^{N,EX}$ and $\Pi_{1,hh}^{N,EX}$ are shown in the same color, since they differ at this density by less than 1\%.  The same color code is applied in Fig.\ \ref{Fig_1}(b) for the pair $(\Pi_{1,ee}^{A,EX}$, $\Pi_{1,hh}^{A,EX})$. These anomalous pairs are, in fact, always formally identical.

\begin{figure}[t]
    \centering
        \includegraphics[trim=0.0cm 0.0cm 0.0cm 0.0cm, clip=true, width=0.5\textwidth]{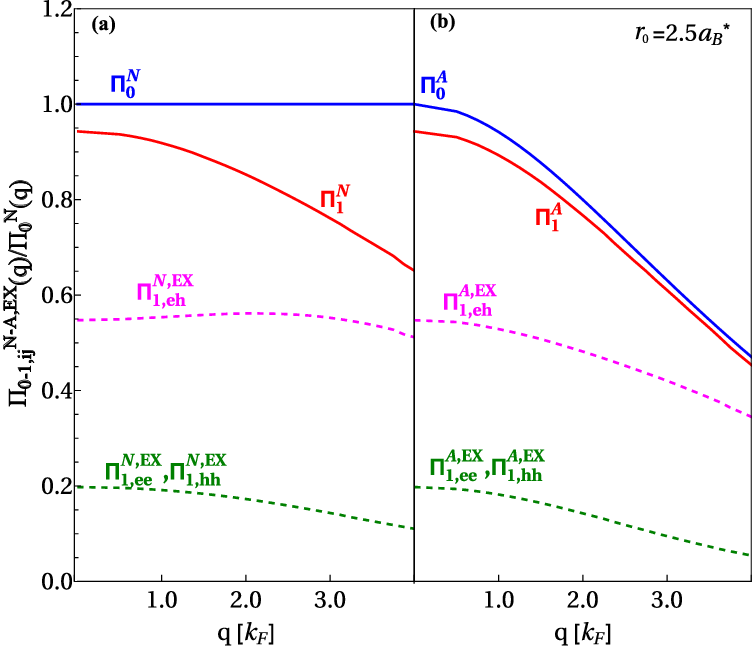}
        \caption{(a) Normal zero-order polarization function $\Pi_0^N$ (Eq.\ \eqref{P0N}, blue solid line).  Normal first-order polarization function $\Pi_1^N$ (red solid line) with its components as in Fig.\ \ref{Fig_1}. (b) Anomalous zero-order polarization function $\Pi_0^A$ (Eq.\ \eqref{P0A}, blue solid line). Anomalous first-order polarization function $\Pi_1^A$ (red solid line) with its components as in Fig.\ \ref{Fig_1}. Interparticle distance $r_0=2.5a_B^*$. All the curves are in unit of $\Pi_0^N(\mathbf{q})$.}
    \label{Fig.2}
\end{figure}

Figure \ref{Fig.2} shows the zero-order and first-order contributions to the normal and anomalous polarization functions at a higher density,  $r_0=2.5$$a_B^*$. At such density, the zero-order terms $\Pi_0^NV^S$ and $\Pi_0^AV^D$ have increased by an order of magnitude with respect to $r_s=10.0$$a_B^*$. Comparing with Fig.\ \ref{Fig_1}, we see that the terms $\Pi_{1,ee}^{N}$, $\Pi_{1,hh}^{N}$ and $\Pi_{1,ee}^{A}$, $\Pi_{1,hh}^{A}$ have all increased respect to $\Pi_0^N$ because the exciton condensate is less strong.  
This is because, with increasing density the system evolves from a BEC toward a BCS regime and the energy gap $\Delta_\mathbf{k}$ becomes weaker than the Fermi energy \cite{Pieri2007}. 
With a smaller energy gap, there are more available low-lying states for the electron-electron and hole-hole scattering channels. 
Figure \ref{Fig.2} shows that the cancellations between the normal and anomalous contributions, both for zero-order and first-order, are restricted to a much smaller range of $\mathbf{q}$ than in Fig.\ \ref{Fig_1}. 

\begin{figure}[b]
    \centering
    \includegraphics[trim=0.0cm 0.0cm 0.0cm 0.0cm, clip=true, width=0.43\textwidth]{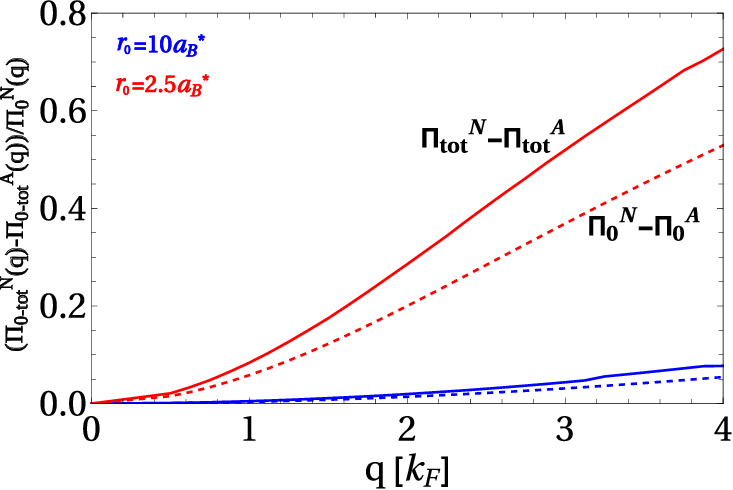}
        \caption{The differences $(\Pi_0^N-\Pi_0^A)$ (dashed lines) and $\Pi_\textrm{tot}^N-\Pi_\textrm{tot}^A\equiv \Pi_0^N+\Pi_1^S-\Pi_0^A-\Pi_1^D$ (solid lines) in units of $\Pi_0^N(q)$.  Red curves  $r_0=2.5a_B^*$ 
        and blue curves $r_0=10a_B^*$. All the curves are in unit of $\Pi_0^N(q)$.}
    \label{Fig.3}
\end{figure}

Figure \ref{Fig.3} then shows the difference to zero-order $(\Pi_0^N-\Pi_0^A)$, and the total difference $(\Pi_\textrm{tot}^N-\Pi_\textrm{tot}^A)$, with $\Pi_\textrm{tot}^N\equiv\Pi_0^N+\Pi_1^S$ and $\Pi_\textrm{tot}^A\equiv\Pi_0^A+\Pi_1^A$.

At a very low density,  $r_0=10a_B^*$, the cancellation at zero-order between the normal and anomalous parts is near complete, and the inclusion of the first-order corrections has little effect. However, at a larger density, $r_0=2.5a_B^*$, the zero-order cancellation is less complete, and the inclusion of the first-order corrections does significantly enhance the difference $(\Pi_\textrm{tot}^N-\Pi_\textrm{tot}^A)$ at large momenta. 

Quantum Monte Carlo simulations \cite{LopezRios2018} and self-consistent zero-order RPA calculations \cite{Pascucci2024} show that for $r_0< 2.5a_B^*$, the screening rapidly strengthens, leading to suppression of the superfluid phase over a short interval of $r_0$.
For this reason, we limit our results to $r_0\geq 2.5a_B^*$ for which the superfluid phase is strong.

\begin{figure}[t]
    \centering
    \includegraphics[trim=0.0cm 0.0cm 0.0cm 0.0cm, clip=true, width=0.5\textwidth]{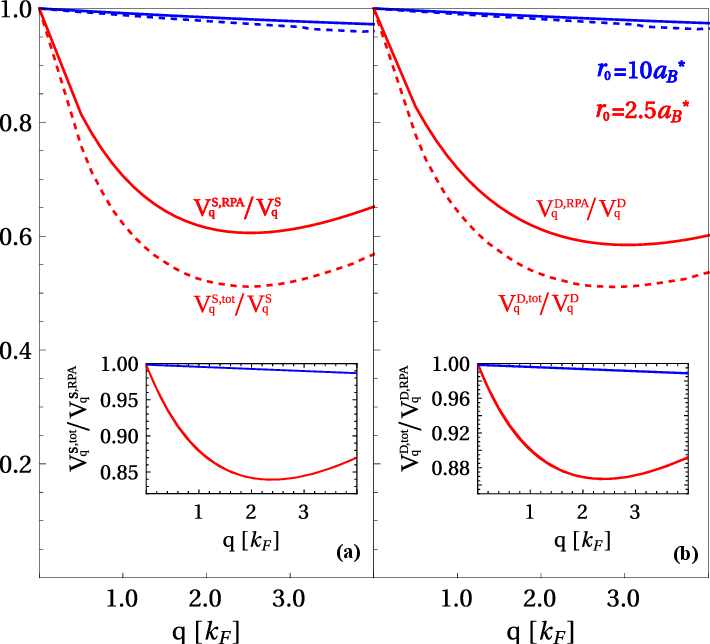}
        \caption{(a) Ratio of the electron-electron screened interaction to the bare Coulomb interaction: solid lines correspond to the zero-order (RPA) result, and dashed lines to the first-order screened interaction. The inset shows the ratio between the zero-order and first-order screened interactions. Results are shown for $r_0=10a_B^*$ (blue) and $r_0=2.5a_B^*$ (red).
(b) Same as in panel (a), but for the electron-hole interaction.}
    \label{Fig.4}
\end{figure}

Figure \ref{Fig.4} shows the extent to which the first-order corrections affect the screened interactions in exciton superfluidity. 
At a low density, $r_0 = 10a_B^*$,  the RPA zero-order screened interactions and the first-order screened interactions - whether attractive or repulsive - closely match the bare Coulomb interaction. The inset shows the ratio of the screening taken to zero-order and taken to first-order, and we see that first-order corrections have little effect.

At higher density, $r_0=2.5a_B^*$, Fig.\ \ref{Fig.4} shows that the zero-order screened interaction is significantly reduced compared to the bare interaction - by $\sim\!\!30$\% at $k_F$, and that including first-order corrections causes a further reduction in the interaction strength. At small momenta, the difference $(\Pi^N-\Pi^A)$ is only weakly affected by first-order corrections, resulting in a correspondingly small change in the screened interaction. Increasing the momentum the difference $(\Pi^N-\Pi^A)$ is more sensitive to first-order corrections.
The insets in Fig.\ \ref{Fig.4} shows that at $\mathbf{k}=2k_F$ the first-order corrections reduce the zero-order RPA interactions by $\sim\!\!12-15$\%. For $k\gg 2k_F$ the increasing difference $(\Pi^N-\Pi^A)$ does not translate into a further reduction of the screened interactions. We find that at large $\mathbf{k}$ values the zero-order RPA and first-order screened interactions merge.  The large momenta states that would participate in the screening are practically empty, and thus the screened interactions recover the bare interactions.

\section{Conclusions}
We have presented for the first time for a 2D bilayer exciton superfluid, a systematic derivation both within the RPA and with all first-order corrections to the RPA included, of the normal and anomalous density polarization functions and of the screened  electron-electron and electron-hole interactions. Consideration of screening is essential in this system because of the long-range Coulomb interactions \cite{Lozovik2012,Perali2013}. 

A crucial feature of our approach is that we do not expand the density polarization functions around the non-interacting Fermi sea ground state, as in normal-state perturbation theory. Instead, we expand the functions around the exciton superfluid mean-field ground state.
This choice has an important consequence: all first-order self-energy insertions cancel exactly against the corresponding terms in the ground state \cite{Nozieres1997}. As a result of this, the only remaining first-order corrections to the polarization functions are exchange terms - the true many-body fluctuations beyond mean-field. 
This structure sharply contrasts with that of Migdal–Eliashberg theory,
where the nature of the electron-phonon interaction gives rise to substantial self-energy corrections, while vertex corrections can be neglected by Migdal’s theorem for superconductors in the adiabatic regime of the electron-phonon coupling.

At low density, we find there is very strong cancellation between the normal and anomalous polarization functions to both zero-order and first-order. As a result, the screened interactions both within the RPA and with the first-order corrections included, are very well approximated by the bare Coulomb interactions. 
At such densities, the first-order polarization functions are dominated by the electron-hole exchange vertex corrections.

At higher densities, the RPA screening starts to become significant because the normal-anomalous cancellation is less complete. With first-order corrections included, the difference between the normal and the anomalous polarization functions at large momenta becomes larger.  However, these corrections have little effect on the screened interactions and hence on the electron-hole pairing strength. 

We conclude that exciton superfluid RPA works well for the screening at least up to the density for the maximum of the superfluid gap energy, the optimal density range in the search for exciton bilayer superfluidity. We recall that at densities higher than this range, the energy gap within RPA becomes rapidly suppressed by screening \cite{Pascucci2024, Nilsson2021}. Including first-order corrections would further reduce the superfluid energy gap at these densities. 
\\
\\
{\bf Acknowledgments}\\
This work was supported by Flemish Research Foundation (FWO-Vl), with contract number 12A3T24N and 1224225N.
A.P. acknowledges partial support by PNRR MUR project PE0000023-NQSTI.

\onecolumngrid
\appendix

\section{Derivation of screened interaction in the exciton superfluid state}
\label{App:screened}

In a two-component system, such as the electron-hole symmetric bilayer considered here, the bare interactions $V(\mathbf{q})$, the screened interaction $ W(\mathbf{q},\omega)$, and the proper polarization $\Pi^*(\mathbf{q},\omega)$ all become 2x2 square matrices. If 
multiplication between matrices is denoted by a dot as in $a\cdot b$ we get  
\begin{equation}
    W(\mathbf{q},\omega)=V(\mathbf{q})+V(\mathbf{q}) \cdot\Pi^*(\mathbf{q},\omega) \cdot W(\mathbf{q},\omega)\ 
    \label{Dyseq2},
\end{equation}
which yields at once the screened interaction matrix,
\begin{equation}
    W(\mathbf{q},\omega)=[{\cal{I}}-{V\cdot \Pi^*(\mathbf{q},\omega)}]^{-1}\cdot V(\mathbf{q}) ,
    \label{Dyseq3}
\end{equation}
  ${\cal{I}}$ being the identity matrix.
The bare interaction matrix 
\begin{align}
    V(\mathbf{q})=
    \begin{pmatrix}
       V^S_\mathbf{q} & V^D_\mathbf{q} \\
       V^D_\mathbf{q} & V^S_\mathbf{q}
    \end{pmatrix}
    \ ,
    \label{VmatrixSF}
\end{align} 
and the proper polarization matrix 
\begin{align}
    \Pi^*(\mathbf{q},\omega)=
    \begin{pmatrix}
       \Pi^{N}(\mathbf{q},\omega) & \Pi^{A}(\mathbf{q},\omega) \\
       \Pi^{A}(\mathbf{q},\omega) & \Pi^{N}(\mathbf{q},\omega) 
    \end{pmatrix} \ ,
    \label{PmatrixSF}
\end{align}
are diagonalized by a rotation of $45^o$ \cite{Sodemann2012} to 
\begin{align}
    \tilde{V}(\mathbf{q})=
    \begin{pmatrix}
       V^S_\mathbf{q}+V^D_\mathbf{q} & 0 \\
       0 & V^S_\mathbf{q}-V^D_\mathbf{q}
    \end{pmatrix}
    \ ,
    \label{VmatrixSF1}
\end{align} 
and  
\begin{align}
    \tilde{\Pi}^*(\mathbf{q},\omega)=
    \begin{pmatrix}
       \Pi^{N}(\mathbf{q},\omega) + \Pi^{A}(\mathbf{q},\omega) & 0\\
       0 &\Pi^{N}(\mathbf{q},\omega) - \Pi^{A}(\mathbf{q},\omega) 
    \end{pmatrix} \ .
    \label{PmatrixSF1}
\end{align}
Using Eq. \eqref{VmatrixSF1} and \eqref{PmatrixSF1} in Eq. \eqref{Dyseq3})
  \begin{align}
 \tilde{W}(\mathbf{q},\omega)=[{\cal{I}}-{\tilde{V}\cdot \tilde{\Pi}^*(\mathbf{q},\omega)}]^{-1}\cdot \tilde{V}(\mathbf{q})  =  
 \begin{pmatrix}
 \frac{V^S_\mathbf{q}+V^D_\mathbf{q}}
 {1-[V^S_\mathbf{q}+V^D_\mathbf{q}]
 [ \Pi^{N}(\mathbf{q},\omega) + \Pi^{A}(\mathbf{q},\omega)]} &   0 \\
0 &  \frac{V^S_\mathbf{q}-V^D_\mathbf{q}}
{1-[V^S_\mathbf{q}-V^D_\mathbf{q}]
 [ \Pi^{N}(\mathbf{q},\omega) - \Pi^{A}(\mathbf{q},\omega)]}  
 \end{pmatrix}.
 \end{align}
A rotation of $-45^o$ yields 
\begin{align}
     W(\mathbf{q},\omega)=
    \begin{pmatrix}
        V^S_{sc}(\mathbf{q},\omega) & V^D_{sc}(\mathbf{q},\omega) \\
        V^D_{sc}(\mathbf{q},\omega) & V^S_{sc}(\mathbf{q},\omega)
    \end{pmatrix}\ ,
\label{WmatrixSF1}
\end{align}
with 
\begin{align}
 V^S_{sc}(\mathbf{q})=&\frac{1}{2} 
  \left[\frac{V^S_\mathbf{q} + V^D_\mathbf{q} } {1-(V^S_\mathbf{q} + V^D_\mathbf{q} )(\Pi^{N}(\mathbf{q},\omega) + \Pi^{A}(\mathbf{q},\omega) )}+
  \frac{V^S_\mathbf{q} - V^D_\mathbf{q} } {1-(V^S_\mathbf{q} - V^D_\mathbf{q} )(\Pi^{N}(\mathbf{q},\omega) - \Pi^{A}(\mathbf{q},\omega)) } \right]  \label{same}
\end{align}
and
\begin{align}
 V^D_{sc}(\mathbf{q})=&\frac{1}{2} 
  \left[\frac{V^S_\mathbf{q} + V^D_\mathbf{q} } {1-(V^S_\mathbf{q} + V^D_\mathbf{q} )(\Pi^{N}(\mathbf{q},\omega) + \Pi^{A}(\mathbf{q},\omega) )}-
  \frac{V^S_\mathbf{q} - V^D_\mathbf{q} } {1-(V^S_\mathbf{q} - V^D_\mathbf{q} )(\Pi^{N}(\mathbf{q},\omega) - \Pi^{A}(\mathbf{q},\omega)) } \right] . 
\label{diff}
\end{align}
A further straightforward manipulation obtains the screened interactions in the static limit $\omega\rightarrow0$ \cite{Pascucci2024}, reported in the main text as Eqs.\ \eqref{eq-uno}-\eqref{eq-tre}. Equations \eqref{same} and \eqref{diff} are  interesting in their own right when investigating numerically the interplay of sums and differences of polarizations and interactions in determining the screened interactions. This has been widely analyzed in the discussion of Fig.\ \ref{Fig.3}.

\section{Zero order polarization functions from Green function techniques}
\label{App:0order}
We calculate the polarization functions at the zero-order from Eqs.\ \eqref{PiSRPA}-\eqref{PiDRPA}, where
\begin{align}
&&\langle\Psi_{BCS}|T[\rho_e(\mathbf{q},\tau)\rho_e(-\mathbf{q},0)]|\Psi_{BCS}\rangle_c=&\sum_{\mathbf{k},\mathbf{k'}}\langle\Psi_{BCS}|T[\sum_\sigma c^\dag_{\mathbf{k}+\mathbf{q},\sigma}(\tau)c_{\mathbf{k},\sigma}(\tau)
\sum_{\sigma'}c^\dag_{\mathbf{k'}-\mathbf{q},\sigma'}(0)c_{\mathbf{k'},\sigma'}(0)]|\Psi_{BCS}\rangle_c \, , \label{Apprhoe}\\
&&\langle\Psi_{BCS}|T[\rho_e(\mathbf{q},\tau)\rho_h(-\mathbf{q},0)]|\Psi_{BCS}\rangle_c=&\sum_{\mathbf{k},\mathbf{k'}}\langle\Psi_{BCS}|T[\sum_{\sigma}c^\dag_{\mathbf{k}+\mathbf{q},\sigma}(\tau)c_{\mathbf{k},\sigma}(\tau)
\sum_{\sigma'}d^\dag_{\mathbf{k'}-\mathbf{q},\sigma'}(0)d_{\mathbf{k'},\sigma'}(0)]|\Psi_{BCS}\rangle_c \, .
\end{align}

We evaluate the time-ordered products using the "extended" Wick theorem \cite{Nozieres1963} considering all possible contractions.
\begin{align}
    \langle\Psi_{BCS}|T[\rho_e(\mathbf{q},\tau)\rho_e(-\mathbf{q},0)]|\Psi_{BCS}\rangle_c&=
    \sum_{\mathbf{k},\mathbf{k'}}\langle T[\sum_{\sigma} c_{\mathbf{k}+\mathbf{q},\sigma}^\dag(\tau) c_{\mathbf{k},\sigma}(\tau)\sum_{\sigma'} c_{\mathbf{k'}-\mathbf{q},\sigma'}^\dag(0) c_{\mathbf{k'},\sigma'}(0)]\rangle \nonumber\\
 &= \sum_{\sigma,\sigma'}  \sum_{\mathbf{k},\mathbf{k'}}\langle T[\wick{ \c1 c_{\mathbf{k}+\mathbf{q},\sigma}^\dag(\tau) \c2 c_{\mathbf{k},\sigma}(\tau) \c2 c_{\mathbf{k'}-\mathbf{q},\sigma'}^\dag(0) \c1 c_{\mathbf{k'},\sigma'}(0)}]\rangle \label{Gwick}
 \\
    &=-\sum_{\sigma,\sigma'}\delta_{\sigma,\sigma'} \sum_{\mathbf{k},\mathbf{k'}}\delta(\mathbf{k'},\mathbf{k}+\mathbf{q})\langle T[c_{\mathbf{k}+\mathbf{q},\sigma}(0) c^{\dag}_{\mathbf{k}+\mathbf{q},\sigma'}(\tau)]\rangle \langle T[c_{\mathbf{k},\sigma}(\tau) c^{\dag}_{\mathbf{k},\sigma'}(0)]\rangle \nonumber \label{T1prod0oree}\\  
    &=-\sum_\sigma\sum_{\mathbf{k}}iG_\sigma(\mathbf{k}+\mathbf{q},-\tau) iG_\sigma(\mathbf{k},\tau)\\
    &=-g\sum_{\mathbf{k}}iG(\mathbf{k}+\mathbf{q},-\tau) iG(\mathbf{k},\tau) \label{Tprod0oree} \ ,
\end{align}
and
\begin{align}
    \langle\Psi_{BCS}|T[\rho_e(\mathbf{q},\tau)\rho_h(-\mathbf{q},0)]|\Psi_{BCS}\rangle_c&=\sum_{\mathbf{k},\mathbf{k'}}\langle T[\sum_{\sigma} c^\dag_{\mathbf{k}+\mathbf{q},\sigma}(\tau)c_{\mathbf{k},\sigma}(\tau)\sum_{\sigma'} d^\dag_{\mathbf{k'}-\mathbf{q},\sigma'}(0)d_{\mathbf{k'},\sigma'}(0)]\rangle_c \nonumber\\
&=\sum_{\sigma,\sigma'}\sum_{\mathbf{k},\mathbf{k'}}\langle  
T[ \wick {\c1  c^\dag_{\mathbf{k}+\mathbf{q},\sigma}(\tau)\c2  c_{\mathbf{k},\sigma}(\tau) \c1 d^\dag_{\mathbf{k'}-\mathbf{q},\sigma'}(0) \c2 d_{\mathbf{k'},\sigma'}}
]\rangle_c \label{Fwick}\\  
    &=\sum_{\sigma,\sigma'}\delta_{\sigma,\sigma'}
    \sum_{\mathbf{k},\mathbf{k'}}\delta(\mathbf{k'},-\mathbf{k})\langle T[d^{\dag}_{-\mathbf{k}-\mathbf{q},\sigma}(0) c^{\dag}_{\mathbf{k}+\mathbf{q},\sigma'}(\tau)]
    \rangle \langle T[c_{\mathbf{k},\sigma}(\tau)d_{-\mathbf{k},\sigma'}(0)]\rangle \nonumber \\ 
    &=\sum_\sigma\sum_{\mathbf{k}}iF_\sigma^*(\mathbf{k}+\mathbf{q},\tau)iF_\sigma(\mathbf{k},\tau)\label{T1prod0oreh}\\
    &=g\sum_{\mathbf{k}}iF^*(\mathbf{k}+\mathbf{q},\tau)iF(\mathbf{k},\tau)\label{Tprod0oreh} \ ,
    \end{align}
where in Eqs.\ \eqref{Gwick}-\eqref{Fwick} are reported only the contractions different from zero that correspond to connected diagrams. In Eqs. \eqref{T1prod0oree}-\eqref{T1prod0oreh} we introduce the normal Green function $iG_\sigma(\mathbf{k},\tau)=\langle T[ c_{\mathbf{k},\sigma}(\tau)c^{\dag}_{\mathbf{k},\sigma}(0)]\rangle$ and the anomalous Green function $iF_\sigma(\mathbf{k},\tau)=\langle T[d_{-\mathbf{k},\sigma}(\tau) c_{\mathbf{k},\sigma}(0)]\rangle$. Because we are considering equal electron and hole layers (same particle density and effective mass) and because the bare Coulomb interactions do not
depend on the spin or valley we can drop the $\sigma$ dependence and introduce in Eqs.\ \eqref{Tprod0oree}-\eqref{Tprod0oreh} the spin-valley degeneracy factor $g$.

The Green functions can be evaluated by rewriting the $c$ ($d$) and $c^\dag$($d^\dag$) with the particle ($\alpha,\alpha^\dag$) and antiparticle ($\beta,\beta^\dag$) Bogoliubov operators \cite{Bogoliubov1958}:
\begin{align}
    c_\mathbf{k}(\tau)=u_\mathbf{k}\alpha_\mathbf{k}(\tau)-v_\mathbf{k}\beta_{-\mathbf{k}}^\dag(\tau)\, ,\\
    c^\dag_\mathbf{k}(\tau)=u_\mathbf{k}\alpha_\mathbf{k}^\dag(\tau)-v_\mathbf{k}\beta_{-\mathbf{k}}(\tau)\, ,\\
    d_\mathbf{k}(\tau)=v_\mathbf{k}\alpha_{-\mathbf{k}}^\dag(\tau) + u_\mathbf{k}\beta_\mathbf{k}(\tau)\, ,\\
    d^\dag_\mathbf{k}(\tau)=v_\mathbf{k}\alpha_{-\mathbf{k}}(\tau) + u_\mathbf{k}\beta^\dag_\mathbf{k}(\tau)\, ,
\end{align}
with $\alpha_\mathbf{\mathbf{k}}(\tau)=e^{-iE_\mathbf{\mathbf{k}}\tau}\alpha_\mathbf{\mathbf{k}}$ and $\beta_\mathbf{\mathbf{k}}(\tau)=e^{-iE_\mathbf{\mathbf{k}}\tau}\beta_\mathbf{\mathbf{k}}$.
The Green functions in Eqs.\ \eqref{Tprod0oree}-\eqref{Tprod0oreh} are:
\begin{align}
    iG(\mathbf{k},\tau-\tau')&=\langle T[c_{\mathbf{k}}(\tau)c^{\dag}_{\mathbf{k}}(\tau')]\rangle \nonumber\\
        &=\theta(\tau-\tau')\langle c_{\mathbf{k}}(\tau)c_{\mathbf{k}}^{\dag}(\tau')\rangle-\theta(\tau'-\tau)\langle c_{\mathbf{k}}^{\dag}(\tau) c_{\mathbf{k}}(\tau)\rangle\nonumber\\
        &=\theta(\tau-\tau')\langle (u_{\mathbf{k}}\alpha_{\mathbf{k}}(\tau)-v_{\mathbf{k}}\beta_{-\mathbf{k}}^\dag(\tau))(u_{\mathbf{k}}\alpha_{\mathbf{k}}^\dag(\tau')-v_{\mathbf{k}}\beta_{-\mathbf{k}}(\tau'))\rangle \nonumber\\
        &\qquad -\theta(\tau'-\tau)\langle (u_{\mathbf{k}}\alpha_{\mathbf{k}}^\dag(\tau')-v_{\mathbf{k}}\beta_{-\mathbf{k}}(\tau'))(u_{\mathbf{k}}\alpha_{\mathbf{k}}(\tau)-v_{\mathbf{k}}\beta_{-\mathbf{k}}^\dag(\tau))\rangle\nonumber\\
        &=\theta(\tau-\tau')u_{\mathbf{k}}^2\langle \alpha_{\mathbf{k}}(\tau)\alpha_{\mathbf{k}}^\dag(\tau')\rangle-\theta(\tau'-\tau)v_{\mathbf{k}}^2\langle \beta_{-\mathbf{k}}(\tau')\beta_{-\mathbf{k}}^\dag(\tau)\rangle\nonumber\\
          &=\theta(\tau-\tau')u^2_{\mathbf{k}}e^{-iE_{\mathbf{k}}(\tau-\tau')/\hbar}-\theta(\tau'-\tau)v^2_{\mathbf{k}}e^{iE_{\mathbf{k}}(\tau-\tau')/\hbar}\, \label{Gnorm} ,
\end{align}
\begin{align}
   iF(\mathbf{k},\tau-\tau')&=\langle T[c_{\mathbf{k}}(\tau)d_{-\mathbf{k}}(\tau')]\rangle \nonumber\\
    &=\langle\theta(\tau-\tau')c_{\mathbf{k}}(\tau)d_{-\mathbf{k}}(\tau')-\theta(\tau'-\tau)d_{-\mathbf{k}}(\tau')c_{\mathbf{k}}(\tau)\rangle \nonumber\\
    &=\theta(\tau-\tau')\langle (u_\mathbf{k} \alpha_{\mathbf{k}}(\tau)-v_\mathbf{k} \beta_{-\mathbf{k}}^\dag(\tau))(v_{-\mathbf{k}} \alpha^\dag_{\mathbf{k}}(\tau')+u_{-\mathbf{k}} \beta_{-\mathbf{k}}(\tau'))\rangle \nonumber\\
    &\quad -\theta(\tau'-\tau)\langle (v_{-\mathbf{k}} \alpha^\dag_{\mathbf{k}}(\tau')+u_{-\mathbf{k}} \beta_{-\mathbf{k}}(\tau')) (u_\mathbf{k} \alpha_{\mathbf{k}}(\tau)-v_\mathbf{k} \beta^\dag_{-\mathbf{k}}(\tau))\rangle \nonumber\\
    &=v_{\mathbf{k}}u_{\mathbf{k}}\left(\theta(\tau-\tau')e^{-iE_{\mathbf{k}}(\tau-\tau')/\hbar}+\theta(\tau'-\tau)e^{iE_{\mathbf{k}}(\tau-\tau')/\hbar}\right)  \, ,
\end{align}

\begin{align}
     iF^*(\mathbf{k},\tau-\tau')&=\langle T[d^\dag_{-\mathbf{k}}(\tau)c^\dag_{\mathbf{k}}(\tau')\rangle\nonumber\\
        &=\langle\theta(\tau-\tau')d_{-\mathbf{k}}^\dag(\tau)c_{\mathbf{k}}^\dag(\tau')-\theta(\tau'-\tau)c_{\mathbf{k}}^\dag(\tau')d_{-\mathbf{k}}^\dag(\tau)\rangle \nonumber\\
    &=\theta(\tau-\tau')\langle (v_{-\mathbf{k}} \alpha_{\mathbf{k}}(\tau)+u_{-\mathbf{k}} \beta_{-\mathbf{k}}^\dag(\tau))(u_\mathbf{k} \alpha_{\mathbf{k}}^\dag(\tau')-v_\mathbf{k} \beta_{-\mathbf{k}}(\tau'))\rangle \nonumber\\
    &\quad -\theta(\tau'-\tau)\langle  (u_\mathbf{k} \alpha_{\mathbf{k}}^\dag(\tau')-v_\mathbf{k} \beta_{-\mathbf{k}}(\tau')) (v_{-\mathbf{k}} \alpha_{\mathbf{k}}(\tau)+u_{-\mathbf{k}} \beta_{-\mathbf{k}}^\dag(\tau))\rangle \nonumber\\   
    &=v_{\mathbf{k}}u_{\mathbf{k}}\left(\theta(\tau-\tau')e^{-iE_{\mathbf{k}}(\tau-\tau')/\hbar}+\theta(\tau'-\tau)e^{iE_{\mathbf{k}}(\tau-\tau')/\hbar}\right) \, \label{F*anom}.
\end{align}

Evidently $F^*(\mathbf{k},\tau-\tau')=F(\mathbf{k},\tau-\tau')$. 
Thus Eq.\ \eqref{Tprod0oree} becomes
\begin{align}
    \langle\Psi_{BCS}|T[\rho_e(\mathbf{q},\tau)\rho_e(-\mathbf{q},0)]|\Psi_{BCS}\rangle_c
    =&-g\sum_{\mathbf{k},\mathbf{k}'}\delta(\mathbf{k}',\mathbf{k}+\mathbf{q})\left(\theta(-\tau)u_{\mathbf{k}+\mathbf{q}}^2e^{iE_{\mathbf{k}+\mathbf{q}}\tau/\hbar}-\theta(\tau)v_{\mathbf{k}+\mathbf{q}}^2e^{-iE_{\mathbf{k}+\mathbf{q}}\tau/\hbar}\right) \nonumber\\
 &\qquad\qquad\qquad\quad\quad \left(\theta(\tau)u_{\mathbf{k}}^2e^{-iE_{\mathbf{k}}\tau/\hbar}-\theta(-\tau)v_{\mathbf{k}}^2e^{iE_{\mathbf{k}}\tau/\hbar}\right) \nonumber \\
 =&g\sum_{\mathbf{k}}\left(\theta(\tau)v_{\mathbf{k}+\mathbf{q}}^2u_{\mathbf{k}}^2e^{-i(E_{\mathbf{k}+\mathbf{q}}+E_{\mathbf{k}})\tau/\hbar}+ \theta(-\tau)u_{\mathbf{k}+\mathbf{q}}^2v_{\mathbf{k}}^2e^{i(E_{\mathbf{k}+\mathbf{q}}+E_{\mathbf{k}})\tau/\hbar}\right) \, ,
\end{align}

and Eq.\ \eqref{Tprod0oreh} becomes
\begin{align}
    \langle\Psi_{BCS}|T[\rho_e(\mathbf{q},\tau)\rho_h(-\mathbf{q},0)]|\Psi_{BCS}\rangle_c=g&\sum_{\mathbf{k},\mathbf{k}'}\delta(\mathbf{k}',-\mathbf{k})v_{\mathbf{k}+\mathbf{q}}u_{\mathbf{k}+\mathbf{q}}\left(-\theta(\tau)e^{-iE_{\mathbf{k}}(\tau)/\hbar}-\theta(-\tau)e^{iE_{\mathbf{k}}(\tau-)/\hbar}\right)\nonumber\\
 &\qquad\qquad\qquad\qquad\quad  +v_{\mathbf{k}}u_{\mathbf{k}}\left(-\theta(-\tau)e^{iE_{\mathbf{k}}\tau/\hbar}-\theta(\tau)e^{-iE_{\mathbf{k}}\tau/\hbar}\right) \nonumber \\
 =&g\sum_{\mathbf{k}}v_{\mathbf{k}+\mathbf{q}}u_{\mathbf{k}+\mathbf{q}}u_{\mathbf{k}}v_{\mathbf{k}} \left(\theta(\tau)e^{-i(E_{\mathbf{k}+\mathbf{q}}+E_{\mathbf{k}})\tau/\hbar}+ \theta(-\tau)e^{i(E_{\mathbf{k}+\mathbf{q}}+E_{\mathbf{k}})\tau/\hbar}\right)\, .
\end{align}

\section{Derivation first-order}
\label{App:1order}
We calculate the first order corrections of the polarization functions rewriting the interaction Hamiltonian $H_{int}(\tau)$ in Eqs.\ \eqref{P1*N}-\eqref{P1*A} as:
\begin{align}
    H'_1(\tau)&=H_1(\tau)-H_{lin}(\tau)\nonumber\\
        &=\frac{1}{A} \sum_{\substack{\mathbf{k},\mathbf{k'},\mathbf{q\ne 0 }\\ \sigma, \sigma'}}\Bigg[\frac{V^{S}_{\mathbf{q}}}{2} \left(c^\dag_{\mathbf{k}+\mathbf{q},\sigma}(\tau)c^\dag_{\mathbf{k'}-\mathbf{q},\sigma'}(\tau)c_{\mathbf{k'},\sigma'}(\tau)c_{\mathbf{k},\sigma}(\tau)+d^\dag_{\mathbf{k}+\mathbf{q},\sigma}(\tau)d^\dag_{\mathbf{k'}-\mathbf{q},\sigma'}(\tau)d_{\mathbf{k'},\sigma'}(\tau)d_{\mathbf{k},\sigma}(\tau) \right)\nonumber\\
        &\qquad\qquad\quad+V^{D}_{\mathbf{q}}c^\dag_{\mathbf{k}+\mathbf{q},\sigma}(\tau)d^\dag_{\mathbf{k'}-\mathbf{q}\sigma'}(\tau)d_{\mathbf{k'},\sigma'}(\tau)c_{\mathbf{k},\sigma}(\tau) \Bigg] \\ \nonumber
        &-\sum_{\mathbf{k},\sigma} \left[ \chi_{\mathbf{k}}\left(c^\dag_{\mathbf{k},\sigma}(\tau)c_{\mathbf{k},\sigma}(\tau)+d^\dag_{\mathbf{k},\sigma}(\tau)d_{\mathbf{k},\sigma}(\tau)\right)+
        \Delta_{\mathbf{k}}c^\dag_{\mathbf{k},\sigma}(\tau)d^\dag_{\mathbf{-k},\sigma}(\tau)+\Delta_{\mathbf{k}}^*d_{\mathbf{-k},\sigma}(\tau)c_{\mathbf{k},\sigma} (\tau) \right]\nonumber\\
        \nonumber
&=H_{int}^{ee}(\tau)+H_{int}^{hh}(\tau)+H_{int}^{eh}(\tau)\ ,
\end{align}
where:
\begin{align}
    H_{int}^{ee}(\tau)&=\frac{1}{2 A} \sum_{\substack{\mathbf{k},\mathbf{k'},\mathbf{q\ne 0}\\\sigma,\sigma'}}
V^{S}_{\mathbf{q}}c^\dag_{\mathbf{k}+\mathbf{q},\sigma}(\tau)
c^\dag_{\mathbf{k'}-\mathbf{q},\sigma'}(\tau)
c_{\mathbf{k'},\sigma'}(\tau)c_{\mathbf{k},\sigma}(\tau)
    -\sum_{\mathbf{k},\sigma}\chi_{\mathbf{k}} c^\dag_{\mathbf{k},\sigma}(\tau)c_{\mathbf{k},\sigma}(\tau)\ ,\\
    H_{int}^{hh}(\tau)&=\frac{1}{A} \sum_{\substack{\mathbf{k},\mathbf{k'},\mathbf{q\ne 0}\\\sigma,\sigma'}}\frac{V^{S}_{\mathbf{q}}}{2} d^\dag_{\mathbf{k}+\mathbf{q},\sigma}(\tau)d^\dag_{\mathbf{k'}-\mathbf{q},\sigma'}(\tau)d_{\mathbf{k'},\sigma'}(\tau)d_{\mathbf{k},\sigma}(\tau)
    -\sum_{\mathbf{k},\sigma}\chi_{\mathbf{k}}\label{eq:H_hh_int}
    d^\dag_{\mathbf{k},\sigma}(\tau)d_{\mathbf{k},\sigma}(\tau)\ ,\\
    H_{int}^{eh}(\tau)&=\frac{1}{A}\!\!\sum_{\substack{\mathbf{k},\mathbf{k'},\mathbf{q \ne 0}\\ \sigma,\sigma'}}V^{D}_{\mathbf{q}} c^\dag_{\mathbf{k}+\mathbf{q},\sigma}(\tau)d^\dag_{\mathbf{k'}-\mathbf{q},\sigma'}(\tau)d_{\mathbf{k'},\sigma'}(\tau)c_{\mathbf{k},\sigma}(\tau) 
    -\sum_{\mathbf{k},\sigma} \left[\Delta_{\mathbf{k}}c^\dag_{\mathbf{k},\sigma}(\tau)d^\dag_{\mathbf{-k},\sigma}(\tau)+\Delta^*_{\mathbf{k}}
    d_{\mathbf{-k},\sigma}(\tau)c_{\mathbf{k},\sigma}(\tau)\right]\ .
    \label{eq:H_eh_int}
\end{align}

The first-order polarization functions are
\begin{align}
    \Pi^{N}_{1}(\mathbf{q},\omega)&=\Pi_{1,ee}^{N}(\mathbf{q},\omega)+\Pi_{1,hh}^{N}(\mathbf{q},\omega)+\Pi_{1,eh}^{N}(\mathbf{q},\omega)\ ,\\
    \Pi^{A}_{1}(\mathbf{q},\omega)&=\Pi_{1,ee}^{ A}(\mathbf{q},\omega)+\Pi_{1,hh}^{ A}(\mathbf{q},\omega)+\Pi_{1,eh}^{ A}(\mathbf{q},\omega)\ ,
    \end{align}
where:
\begin{align}
    \Pi_{1,ij}^{N}(\mathbf{q},\omega)&=-\frac{1}{A\hbar^2}\lim_{\eta\rightarrow0}\iint d\tau d\tau' e^{i(\omega\tau+i\eta|\tau|)}\langle T[\rho_e(\mathbf{q},\tau)\rho_e(-\mathbf{q},0)H_{int}^{ij}(\tau')]\rangle_c\ \label{P1N},\\
        \Pi_{1,ij}^{A}(\mathbf{q},\omega)&=-\frac{1}{A\hbar^2}\lim_{\eta\rightarrow0}\iint d\tau d\tau' e^{i(\omega\tau+i\eta|\tau|)}\langle T[\rho_e(\mathbf{q},\tau)\rho_h(-\mathbf{q},0)H_{int}^{ij}(\tau')]\rangle_c \ \label{P1A}\ .
\end{align}
Note that $\Pi_{1.ee}^{ A}(\mathbf{q},\omega)=\Pi_{1.hh}^{ A}(\mathbf{q},\omega)$ because of the symmetry of the system.
We explicit the $T$ product of $\Pi_{1,ee}^{N}$ (Eq.\ \eqref{P1N}):
\begin{align}
T[\rho_e(\mathbf{q},\!\tau)\rho_e(-\mathbf{q},0)H_{int}^{ee}(\tau')]\!
&= 
\label{eq:Pi_ee_int_TP}
\sum_{\substack{\mathbf{k},\mathbf{k}' \\\gamma,\gamma'}}T\! \left\{ 
c_{\mathbf{k}+\mathbf{q},\gamma}^\dag(\tau) c_{\mathbf{k},\gamma}(\tau)c_{\mathbf{k'}-\mathbf{q},\gamma'}^\dag(0)c_{\mathbf{k'},\gamma'}(0)\! \right.\nonumber
\\
& \left.\left[
\frac{1}{2 A}\! \sum_{\substack{\mathbf{p},\mathbf{p'},\mathbf{s\ne 0}\\\sigma,\sigma'}}
V^{S}_{\mathbf{q}}c^\dag_{\mathbf{p}+\mathbf{s},\sigma}(\tau')
c^\dag_{\mathbf{p'}-\mathbf{s},\sigma'}(\tau')
c_{\mathbf{p'},\sigma'}(\tau')c_{\mathbf{p},\sigma}(\tau')
\!-\!\sum_{\mathbf{p},\sigma}\chi_{\mathbf{p}} 
c^\dag_{\mathbf{p},\sigma}(\tau')c_{\mathbf{p},\sigma}(\tau')\ \right]
\right\} .
\end{align}
We evaluate the time-ordered products using the Wick theorem. The fully contracted connected terms are: (i) non equal-time contractions, the exchange (EX) terms; (ii) the equal time contractions with non-zero exchanged momentum, the self-energy (SE) terms; (iii) the equal time contractions with zero exchanged momentum, Hartree direct terms. These latter they all cancel out between each other because of charge neutrality of the system.
Let's focus on the term in the square brackets in Eq.~\eqref{eq:Pi_ee_int_TP} and consider the equal time contractions with non-zero exchanged momentum:
\begin{align}
\nonumber
& \frac{1}{2 A} \sum_{\substack{\mathbf{p},\mathbf{p'},\mathbf{s\ne 0}\\\sigma,\sigma'}}  
V^{S}_{\mathbf{s}}c^\dag_{\mathbf{p}+\mathbf{s},\sigma}(\tau)
c^\dag_{\mathbf{p'}-\mathbf{s},\sigma'}(\tau)
c_{\mathbf{p'},\sigma'}(\tau)c_{\mathbf{p},\sigma}(\tau)
-\sum_{\mathbf{p},\sigma}\chi_{\mathbf{p}} 
c^\dag_{\mathbf{p},\sigma}(\tau)c_{\mathbf{p},\sigma}(\tau)\  \\
\nonumber
&=\frac{1}{2A} \sum_{\substack{\mathbf{p},\mathbf{p'},\mathbf{q\ne 0}\\\sigma,\sigma'}}
V^{S}_{\mathbf{s}}
\left[\wick{\c1 c^\dag_{\mathbf{p}+\mathbf{s},\sigma}(\tau)
 c^\dag_{\mathbf{p'}-\mathbf{s},\sigma'}(\tau)
\c1 c_{\mathbf{p'},\sigma'}(\tau)c_{\mathbf{p},\sigma}(\tau)+
c^\dag_{\mathbf{p}+\mathbf{s},\sigma}(\tau)
\c2 c^\dag_{\mathbf{p'}-\mathbf{s},\sigma'}(\tau)
c_{\mathbf{p'},\sigma'}(\tau)
\c2 c_{\mathbf{p},\sigma}(\tau)
}\right]
-\sum_{\mathbf{p},\sigma}\chi_{\mathbf{p}} 
c^\dag_{\mathbf{p},\sigma}(\tau)c_{\mathbf{p},\sigma}(\tau)\   
\\
\nonumber
&= -\frac{1}{2 A}
\sum_{\substack{\mathbf{p},\mathbf{p}',\mathbf{s} \ne 0 \\\sigma,\sigma'}}
\delta(\mathbf{p}',\mathbf{p}+\mathbf{s}) \delta_{\sigma,\sigma'}
V^{S}_{\mathbf{s}}
\left[v_{\mathbf{p}+\mathbf{s}}^2 
c_{\mathbf{p},\sigma}^\dag\!(\tau')  
c_{\mathbf{p},\sigma}\!(\tau')+v_{\mathbf{p}-\mathbf{s}}^2 c_{\mathbf{p},\sigma}^\dag\!(\tau')
c_{\mathbf{p},\sigma}\!(\tau')\right]-\sum_{\mathbf{p},\sigma}
\chi_{\mathbf{p}} 
c^\dag_{\mathbf{p},\sigma}(\tau')c_{\mathbf{p},\sigma}(\tau')\ 
\\
&=\sum_{\mathbf{p},\sigma}
c^\dag_{\mathbf{p},\sigma}(\tau)c_{\mathbf{p},\sigma}(\tau)
\left[
-\chi_p -\frac{1}{A}\sum_{\mathbf{s}\ne 0} V^{S}_{\mathbf{s}} 
v_{\mathbf{p}+\mathbf{s}}^2
\right] =0\ .
\end{align} 
Thanks to Eq.~\eqref{eq:hf_nrg} this contribution is null, and the same happens with the the hole-hole term, Eq.\ \eqref{eq:H_hh_int}.
Similarly, the self-energy terms coming from the insertion of
\eqref{eq:H_eh_int} are null thanks to Eq.~\eqref{eq:gap_eq}.
The normal self-energy contributions in the interaction terms of $H_1$ cancel out with on in the ground-state $H_{lin}$. 
The same happens with the self-energy contributions in the anomalous terms. The reason it that the ground-state described by $H_{lin}$ already contains the Hartree-Fock and the superfluid gap energy. 
Thus, the normal and anomalous first-order corrections consist only of exchange contributions.
In particular, for each term, the exchange contributions appear twice, so from now on, we multiply each term by $2$.
Because we are considering equal electron and hole layers and because the bare Coulomb interactions do not depend on the spin or valley, for simplicity we can drop the $\sigma$ dependence and introduce the spin-valley degeneracy factor $g$.

The exchange terms from Eq.\ \eqref{P1N} read:
\begin{align}
    \Pi^{N,EX}_{1,ee}(\mathbf{q})&=\frac{g}{\hbar^2A^2}\ \lim_{\eta,\omega\to 0}\sum_{\mathbf{k},\mathbf{p}}\iint d\tau d\tau' e^{i(\omega\tau+i\eta|\tau|)} iG(\mathbf{k}+\mathbf{q},\tau'-\tau)iG(\mathbf{k},\tau-\tau')iG(\mathbf{p}+\mathbf{q},-\tau')iG(\mathbf{p},\tau')V^{S}_{\mathbf{k}-\mathbf{p}} \nonumber\\
    &=\frac{g}{A^2\hbar^2}\ \lim_{\eta,\omega\to 0} \sum_{\mathbf{k},\mathbf{p}}\iint d\tau d\tau' e^{i(\omega\tau+i\eta|\tau|)}\left(\theta(\tau'-\tau)u_{\mathbf{k}+\mathbf{q}}^2e^{iE_{\mathbf{k}+\mathbf{q}}(\tau'-\tau)/\hbar}-\theta(\tau-\tau')v_{\mathbf{k}+\mathbf{q}}^2e^{-iE_{\mathbf{k}+\mathbf{q}}(\tau'-\tau)/\hbar}\right) \nonumber\\
    &\qquad \qquad \qquad \qquad\qquad \qquad \qquad \qquad \qquad \qquad  \left(\theta(\tau-\tau')u_{\mathbf{k}}^2e^{iE_{\mathbf{k}}(\tau-\tau')/\hbar}-\theta(\tau'-\tau)v_{\mathbf{k}}^2e^{-iE_{\mathbf{k}}(\tau-\tau')/\hbar}\right) \nonumber\\
    &\quad  \left(\theta(-\tau')u_{\mathbf{p}+\mathbf{q}}^2e^{-iE_{\mathbf{p}+\mathbf{q}}\tau'/\hbar}-\theta(\tau')v_{\mathbf{p}+\mathbf{q}}^2e^{iE_{\mathbf{p}+\mathbf{q}}\tau'/\hbar}\right)\left(\theta(\tau')u_{\mathbf{p}}^2e^{iE_{\mathbf{p}}\tau'/\hbar}-\theta(-\tau')v_{\mathbf{p}}^2e^{-iE_{\mathbf{p}}\tau'/\hbar}\right) V^{S}_{\mathbf{k}-\mathbf{p}} \nonumber \\
    &=\frac{g}{A^2\hbar^2}\ \lim_{\eta,\omega\to 0}\sum_{\mathbf{k},\mathbf{p}}\iint d\tau d\tau' e^{i(\omega\tau+i\eta|\tau|)}\nonumber\\
    &\qquad \qquad \qquad \qquad\qquad   \left(\theta(\tau'-\tau)u_{\mathbf{k}+\mathbf{q}}^2v_\mathbf{k}^2e^{-i(E_{\mathbf{k}+\mathbf{q}}+E_{\mathbf{k}})(\tau-\tau')/\hbar}+\theta(\tau-\tau')v_{\mathbf{k}+\mathbf{q}}^2u_\mathbf{k}^2e^{i(E_{\mathbf{k}+\mathbf{q}}+E_{\mathbf{k}})(\tau-\tau')/\hbar}\right) \nonumber\\
    &\qquad \qquad \qquad \qquad\qquad   \left(\theta(-\tau')u_{\mathbf{p}+\mathbf{q}}^2v_{\mathbf{p}}^2e^{-i(E_{\mathbf{p}+\mathbf{q}}+E_{\mathbf{p}})\tau'/\hbar}+\theta(\tau')v_{\mathbf{p}+\mathbf{q}}^2u_\mathbf{p}^2e^{i(E_{\mathbf{p}+\mathbf{q}}+E_{\mathbf{p}})\tau' /\hbar}\right) V^{S}_{\mathbf{k}-\mathbf{p}} \nonumber\\
    \Pi^{N,EX}_{1,ee}(\mathbf{q})&=-\frac{g}{A^2}\sum_{\mathbf{k},\mathbf{p}}\frac{(u_{\mathbf{p}+\mathbf{q}}^2v_{\mathbf{p}}^2+u_{\mathbf{p}}^2v_{\mathbf{p}+\mathbf{q}}^2)(u_{\mathbf{k}+\mathbf{q}}^2v_{\mathbf{k}}^2+u_{\mathbf{k}}^2v_{\mathbf{k}+\mathbf{q}}^2)}{\left(E_{\mathbf{p}}+E_{\mathbf{p}+\mathbf{q}}\right)\left(E_{\mathbf{k}}+E_{\mathbf{k}+\mathbf{q}}\right)}V^{S}_{\mathbf{k}-\mathbf{p}}\ ,
\end{align}
and
\begin{align}
     \Pi_{1,hh}^{N,EX}(\mathbf{q})&=\frac{g}{\hbar^2A^2}\sum_{\mathbf{k},\mathbf{p}} \lim_{\eta,\omega\to 0}\iint d\tau d\tau' e^{i(\omega\tau+i\eta|\tau|)}iF^*(\mathbf{k}+\mathbf{q},\tau-\tau')iF(\mathbf{p}+\mathbf{q},\tau')iF^*(\mathbf{p},-\tau')iF(\mathbf{k},\tau'-\tau)V^{S}_{\mathbf{k}-\mathbf{p}}\nonumber\\
     &=-4\frac{g}{A^2}\sum_{\mathbf{k},\mathbf{p}}\frac{u_\mathbf{k}v_\mathbf{k}u_{\mathbf{k}+\mathbf{q}}v_{\mathbf{k}+\mathbf{q}}u_{\mathbf{p}}v_{\mathbf{p}}u_{\mathbf{p}+\mathbf{q}}v_{\mathbf{p}+\mathbf{q}}}{(E_{\mathbf{k}}+E_{\mathbf{k}+\mathbf{q}})(E_{\mathbf{p}}+E_{\mathbf{p}+\mathbf{q}})}V^{S}_{\mathbf{k}-\mathbf{p}} \ ,
\end{align}

\begin{align}
    \Pi^{N,EX}_{1,eh}(\mathbf{q})&=-\frac{2g}{A^2\hbar^2} \lim_{\eta,\omega\to 0}\sum_{\mathbf{k},\mathbf{p}}\iint d\tau d\tau' e^{i(\omega\tau+i\eta|\tau|)} iG(\mathbf{k}+\mathbf{q},\tau'-\tau)iG(\mathbf{p}+\mathbf{q},-\tau')iF^*(\mathbf{p},-\tau')iF(\mathbf{k},\tau'-\tau)V^{D}_{\mathbf{k}-\mathbf{p}}\nonumber\\
    &=\frac{2g}{A^2}\sum_{\mathbf{k},\mathbf{p}}\frac{u_\mathbf{k}v_\mathbf{k}u_\mathbf{p} v_\mathbf{p}(v_{\mathbf{p}+\mathbf{q}}^2-u_{\mathbf{p}+\mathbf{q}}^2)(v_{\mathbf{k}+\mathbf{q}}^2-u_{\mathbf{k}+\mathbf{q}}^2)}{(E_{\mathbf{p}}+E_{\mathbf{p}+\mathbf{q}})(E_{\mathbf{k}}+E_{\mathbf{k}+\mathbf{q}})}V^{D}_{\mathbf{k}-\mathbf{p}}\ .
\end{align}

The exchange corrections from Eqs.\ \eqref{P1A}  are:

\begin{align}
    \Pi^{A,EX}_{1,ee}(\mathbf{q})&=\Pi^{A,EX}_{1,hh}(\mathbf{q})\nonumber\\
    &=-\frac{g}{A^2\hbar^2} \lim_{\eta,\omega\to 0}\sum_{\mathbf{k},\mathbf{p}}\iint d\tau d\tau' e^{i(\omega\tau+i\eta|\tau|)} iG(\mathbf{k}+\mathbf{q},\tau'-\tau)iF^*(\mathbf{p}+\mathbf{q},\tau')iF(\mathbf{p},-\tau')iG(\mathbf{k},\tau-\tau')V^{S}_{\mathbf{p}-\mathbf{k}}\nonumber\\
    &=-\frac{g}{A^2}\sum_{\mathbf{k},\mathbf{p}}\frac{v_\mathbf{k}^2u_{\mathbf{k}+\mathbf{q}}^2+u_\mathbf{k}^2v_{\mathbf{k}+\mathbf{q}}^2}{E_{\mathbf{k}}+E_{\mathbf{k}+\mathbf{q}}}\frac{2v_{\mathbf{p}+\mathbf{q}}u_{\mathbf{p}+\mathbf{q}}v_{\mathbf{p}}u_{\mathbf{p}}}{E_{\mathbf{p}}+E_{\mathbf{p}+\mathbf{q}}}V^{S}_{\mathbf{k}-\mathbf{p}}\ ,
\end{align}

\begin{align}
    \Pi^{A,EX}_{1,eh}(\mathbf{q})&=\frac{2g}{A^2\hbar^2} \lim_{\eta,\omega\to 0}\sum_{\mathbf{k},\mathbf{p}}\iint d\tau d\tau' e^{i(\omega\tau+i\eta|\tau|)} iG(\mathbf{k}+\mathbf{q},\tau'-\tau)iF^*(\mathbf{p}+\mathbf{q},\tau')iG^*(\mathbf{p},\tau')iF(\mathbf{k},\tau'-\tau)V^{D}_{\mathbf{k}-\mathbf{p}}\nonumber\\
    &=-\frac{2g}{A^2}\sum_{\mathbf{k},\mathbf{p}}\frac{u_\mathbf{k}v_\mathbf{k}u_{\mathbf{p}+\mathbf{q}}v_{\mathbf{p}+\mathbf{q}}(v_{\mathbf{k}+\mathbf{q}}^2-u_{\mathbf{k}+\mathbf{q}}^2)(u_{\mathbf{p}}^2-v_{\mathbf{p}}^2)}{(E_{\mathbf{k}}+E_{\mathbf{k}+\mathbf{q}})(E_{\mathbf{p}}+E_{\mathbf{p}+\mathbf{q}})}V^{D}_{\mathbf{k}-\mathbf{p}}\ .
\end{align}

\twocolumngrid

%

\end{document}